\documentclass[%
 reprint,
superscriptaddress,
 amsmath,amssymb,
 aps,
pre,
]{revtex4-1}
\usepackage[normalem]{ulem}
\usepackage{graphicx, xcolor}
\usepackage[caption=false]{subfig}
\usepackage{dcolumn}
\usepackage{bm}

\begin{document}

\preprint{APS/123-QED}

\title{Effect of anisotropy on the formation of active particle films}

\author{T. C. Rebocho}
\affiliation{Departamento de F\'{\i}sica, Faculdade de Ci\^{e}ncias,Universidade de Lisboa, 1749-016 Lisboa, Portugal}
\affiliation{Centro de F\'{\i}sica Te\'{o}rica e Computacional, Faculdade de Ci\^{e}ncias, Universidade de Lisboa, 1749-016 Lisboa, Portugal}

\author{M. Tasinkevych}
\affiliation{Departamento de F\'{\i}sica, Faculdade de Ci\^{e}ncias,Universidade de Lisboa, 1749-016 Lisboa, Portugal}
\affiliation{Centro de F\'{\i}sica Te\'{o}rica e Computacional, Faculdade de Ci\^{e}ncias, Universidade de Lisboa, 1749-016 Lisboa, Portugal}
\affiliation{SOFT Group, School of Science and Technology, Nottingham Trent University, Clifton Lane, Nottingham NG11 8NS, UK}

\author{C. S. Dias}
\affiliation{Departamento de F\'{\i}sica, Faculdade de Ci\^{e}ncias,Universidade de Lisboa, 1749-016 Lisboa, Portugal}
\affiliation{Centro de F\'{\i}sica Te\'{o}rica e Computacional, Faculdade de Ci\^{e}ncias, Universidade de Lisboa, 1749-016 Lisboa, Portugal}

\email{csdias@fc.ul.pt}




\date{\today}

\begin{abstract}
  Active colloids belong to a class of non-equilibrium systems where energy uptake, conversion and dissipation occurs at the level of individual colloidal particles, which can lead to particles' self-propelled motion and surprising collective behaviour. Examples include coexistence of vapour- and liquid-like steady states for active particles with repulsive interactions only, phenomena known as motility induced phase transition. Similarly to motile unicellular organisms, active colloids tend to accumulate at confining surfaces forming dense adsorbed films. In this work, we study the structure and dynamics of aggregates of self-propelled particle near confining solid surfaces, focusing on the effects of the particle anisotropic interactions. We performed Langevin dynamics simulations of two complementary models for active particles: ellipsoidal particles interacting through Gay–Berne potential, and rod-like particles composed of several repulsive Lennard-Jones beads. We observe a non-monotonic behavior of the structure of clusters formed along the confining surface as a function of the particle aspect ratio, with a film spreading when particles are near spherical, compact clusters with hedgehog-like particle orientation for more elongated active particles, and a complex dynamical behavior for intermediate aspect ratio. The stabilization time of cluster formation along the confining surface also displays a non-monotonic dependence on the aspect ratio, with a local minimum at intermediate values. Additionally, we demonstrate that the hedgehog-like aggregates formed by Gay-Berne ellipsoids exhibit higher structural stability as compared to the ones formed by purely repulsive active rods, which are stable due to the particle activity only.

\end{abstract}

\maketitle


    \section{Introduction}

Active particles belong to non-equilibrium systems with a persistent local entropy production, which violates detailed balance \cite{Neta2021-wetting-solid-surface, Das2018-active-brownian-particles,VanDamme2019-interparticle-torques-suppress-MIPS}. These active particles persistently consume the medium free energy to power their self-propelled motion along a certain direction \cite{Ramaswamy2010-statistics-of-active-matter, Wensink2008-aggregation-colloidal-rods-near-walls, Bar2020-insights-active-matter, Das2018-active-brownian-particles,Huber2018-ordered-states-active-matter,VanDamme2019-interparticle-torques-suppress-MIPS, Lowen2020-Brownian-to-Langevin}. Particle self-propulsion gives rise to novel types of collective behavior such as the coexistence of vapour- and liquid-like steady states for active particles with only repulsive interactions \cite{Neta2021-wetting-solid-surface, Grosmann2020-particle-field-collective-motion-active-matter,Fily2012-athermal-phase-separation,Redner2013-structure-dynamics-phases,Levis2014-clustering-dynamics}, a phenomenon known as motility-induced phase separation (MIPS) \cite{Cates2015-MIPS}.

The interest in the collective behavior of active particles has been driven by numerous applications in theoretical biology \cite{Kolomeisky2007-theorist-perspective}, nonlinear physics, synthetic self-propelled particles development, and pollution remediation systems \cite{Neta2021-wetting-solid-surface, Hernandez-Ortiz2005-transport-confined-swimming-particles, Grosmann2020-particle-field-collective-motion-active-matter, Bar2020-insights-active-matter, Abkenar2013-penetrable-self-propelled, Das2018-active-brownian-particles,Wensink2012-from-swarming-to-turbulence,VanDamme2019-interparticle-torques-suppress-MIPS}. Theoretical and numerical studies of active particles help to understand the behavior of flocks \cite{Mora2016-bird-flocks}, schools \cite{Reuter1994-fish-schools}, herds \cite{Gueron1996-herds-dynamics}, cell aggregates \cite{Acemel2018-bacterial-biofilms}, and artificial microswimmers (like Janus particles) \cite{Jiang2010-janus-particles}.

Significant effort has been directed towards understanding the effects of particle anisotropy on emerging collective behavior of elongated active particles
\cite{Wensink2012-from-swarming-to-turbulence, Abkenar2013-penetrable-self-propelled, Weitz2015-phase-separated-state, VanDamme2019-interparticle-torques-suppress-MIPS, Bar2020-insights-active-matter, Jayaram2020-scalar-to-polar-active-matter, ginelli-properties_rods-2010, Wittkowski-density_colloidal_particles-2011, Grosmann2020-particle-field-collective-motion-active-matter}. In contrast to self-propelled discs or spheres, rod-like active particles exhibit
a zoo of different emergent non-equilibrium states such as motile clusters, turbulence, and lanes \cite{Wensink2012-from-swarming-to-turbulence,Grosmann2020-particle-field-collective-motion-active-matter}. Additionally, excluded volume torques acting between active rods may suppress MIPS, provided the particle aspect ratio
is large enough \cite{VanDamme2019-interparticle-torques-suppress-MIPS, Bar2020-insights-active-matter,Jayaram2020-scalar-to-polar-active-matter, Weitz2015-phase-separated-state}. The occurrence of MIPS is driven by collision-induced particle slowing down in crowded regions and the accumulation of active particles in those regions where they move slower \cite{Schnitzer1993-theory-random-wals}. This positive feedback mechanism leads to the growth of denser particle domains and ultimately results in the phase separation \cite{Cates2015-MIPS}.    
On the other hand, the mentioned above slowing down is related to the duration of particle collisions \cite{Fily2014-freezing-phase-separation},  which can be shorten by the excluded volume torques between elongated particles, thereby pushing MIPS to higher values of the packing fraction and completely suppressing it for the large enough aspect ratio  \cite{VanDamme2019-interparticle-torques-suppress-MIPS}.

In the last few years, an increased interest has emerged over the effect of confinement on the collective behavior of active particles \cite{Wensink2008-aggregation-colloidal-rods-near-walls, Elgeti2009-rods-near-surface,Neta2021-wetting-solid-surface,Sepulveda2017-wetting-transitions,Sepulveda2018-wetting-transitions,Wioland-bacterial_suspension-2013, Lushi-swimming_bacteria-2014, Wioland2016, Bar2020-insights-active-matter, Costanzo-bacteria_microchannel-2012, kaiser-capturing_self_propelled-2013, deblais-boundaries_collective-2018, Elgeti2016, Zargar2009, Xiao2018, Wensink2013, Yu2016, Wang2018}. Similar to bacteria \cite{Picioreanu2004-Biofilms}, active particles have a tendency to build up at confining surfaces forming dense adsorbed films \cite{Neta2021-wetting-solid-surface,Sepulveda2017-wetting-transitions,Sepulveda2018-wetting-transitions}. Moreover, the thickness of the adsorbed films was shown to grow, with a signature of divergence, as the system is brought towards the MIPS coexistence curve from the vapour side \cite{Neta2021-wetting-solid-surface}. For rod-like active particles confined in slit-like channels very rich collective behavior was reported in \cite{Wensink2008-aggregation-colloidal-rods-near-walls}, including formation at the channel surfaces 
of compact immobile clusters with hedgehog-like orientation of active rods. These clusters were observed at intermediate times. At later times the hedgehog aggregates dissolved into quasi-planar surface sliding films with homeotropic particle orientation. Ref.~\cite{Wensink2008-aggregation-colloidal-rods-near-walls} considered only rigid rods. However, it is known that anisotropic attractive interactions can strongly affect growth of films of passive colloids \cite{Yunker2013, Dias2018, Dias2014, Araujo2015}. Additionally, the effects of aspect ratio of active particles on the structure and dynamics of films and clusters formed at confining surfaces remains largely unexplored.       

Here we report a simulation study of two complementary models for elongated active particles. The first model of ellipsoidal particles interacting via Gay-Berne potential, incorporates both the effect of geometry and attractive interaction anisotropy. The second model of rod-like particles, considers only the geometrical anisotropy, where the particles are composed of a given number of repulsive Lennard-Jones beads. We find that the adsorption dynamics and the steady state particle configurations depend in a non-monotonic way on the particle anisotropy. We rationalize this findings in terms of a shape-dependent coefficient of rotational diffusion, and of the surface cluster dynamics. In the next section we provide more details about the used models and the numerical approach. In Sec.~\ref{results} we discuss our results, focusing on the dependence of the adsorption dynamics and the structure of the adsorbed films or aggregates on the particle aspect ratio. Finally, in the last section we present our conclusions.

    \begin{figure}[tb]
\includegraphics[width=1.0\columnwidth]{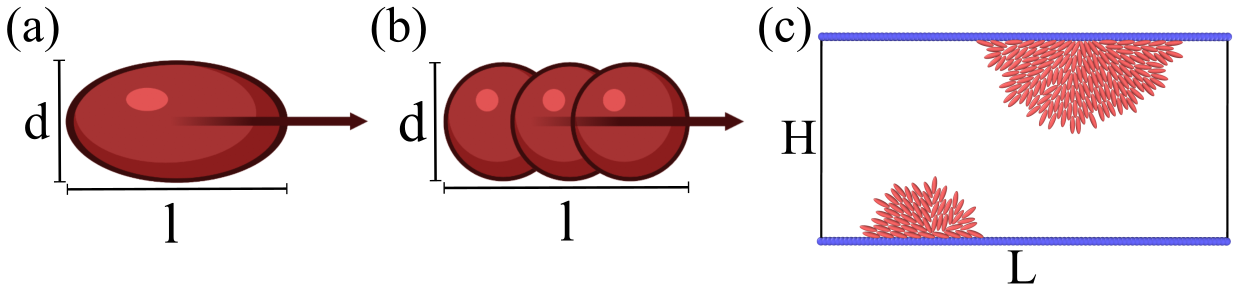}
\caption{Schematic representation of the particle and box geometry: (a) an ellipsoid with length $l$ and width $d$, and (b) a linear chain of length $l$ of 3 disks of diameter $d$. The long axis of the particles defines the self-propulsion direction. (c) Schematic representation of the channel geometry with the length $L$ (along the $x-$axis) and the width $H$ (along the $y-$axis). The slit walls (blue) are placed at $y=0$ and at $y=L$. For the sake of computational efficiency only one chain is modeled explicitly and periodic boundary conditions are also applied in the $y-$direction.}
  \label{fig:schema-ellipsoid-and-rod}
\end{figure}

\section{Model}
\label{methods}

We consider $N$ active Brownian particles with an aspect ratio $\kappa=l/d$ (ellipsoids or linear chains of beads,
see Figure \ref{fig:schema-ellipsoid-and-rod}), where $l$ is the length and $d$ is the width of the particles. We keep the projected area of each particle constant such that $A=\pi d_0^2 /4=\pi ld /4$,
where $d_0$ is the diameter of a particle at $\kappa = 1$ (sphere). 
Each particle
has an intrinsic self-propulsion force of intensity ${F_A}$ directed along the particle long axis \cite{tenHagen2015}. Particles are confined
to a two-dimensional (2D) slit geometry of width $H$ and length $L$ (see Fig.\ref{fig:schema-ellipsoid-and-rod}(c)). The slit walls are modeled 
as fixed linear chains of spherical particles with the diameter $d$.
In the direction of the slit walls we consider periodic boundary conditions.

To resolve the active particle trajectories, we integrate
their equations of motion, using a velocity Verlet scheme implemented
in the Large-scale Atomic/Molecular Massively Parallel Simulator
(LAMMPS) \cite{PLIMPTON1995}. Specifically, the active particles dynamics follows the Langevin equations for translational motion in the $(x,y)$ plane, 
\begin{equation}
 m\dot{\vec{v}}(t)=-\nabla_{\vec{r}} U(\vec{r})-\gamma_t\vec{v}(t)+\sqrt{2\gamma_tk_BT}\vec{\xi}(t)+F_A\hat{v}(t), \label{eq.trans_Langevin_dynamics}
\end{equation}
and rotations around the $z-$ axis (perpendicular to the $(x,y)$ plane), as
\begin{equation}
 I\dot{\omega}(t)=-\nabla_{\vec{r}} U(\theta)-\gamma_r\omega(t)+\sqrt{2\gamma_rk_BT}\xi(t).\label{eq.rot_Langevin_dynamics}
\end{equation}
$\vec{v}$ and $\omega$ 
are the translational and angular velocities, $\gamma_t=\frac{m}{\tau_t}$ and $\gamma_r=\frac{I}{\tau_r}$ are the translational and rotational damping coefficient, $\hat{v}=\vec{v}/\lvert\lvert\vec{v}\rvert\rvert$, $F_A$ the strength of the propulsion force,  $\tau_t$ and $\tau_r$ the translational and rotational damping times, $T$ the temperature, $m$ and $I$ are the mass and the moment of inertia of the particle, and $U$ is the potential energy encoding the interactions with the other particles (including those of the slit walls). $\xi_{t}(t)$  and $\xi_{r}(t)$ are stochastic terms that fulfill the fluctuation-dissipation theorem. We consider values of constants such as the persistence length, $l_p=v/D_r>H$, where $v$ is the particles propulsion velocity and $D_r$ the rotational diffusion coefficient.

\begin{figure}[tb]
        \includegraphics[width=0.9\columnwidth]{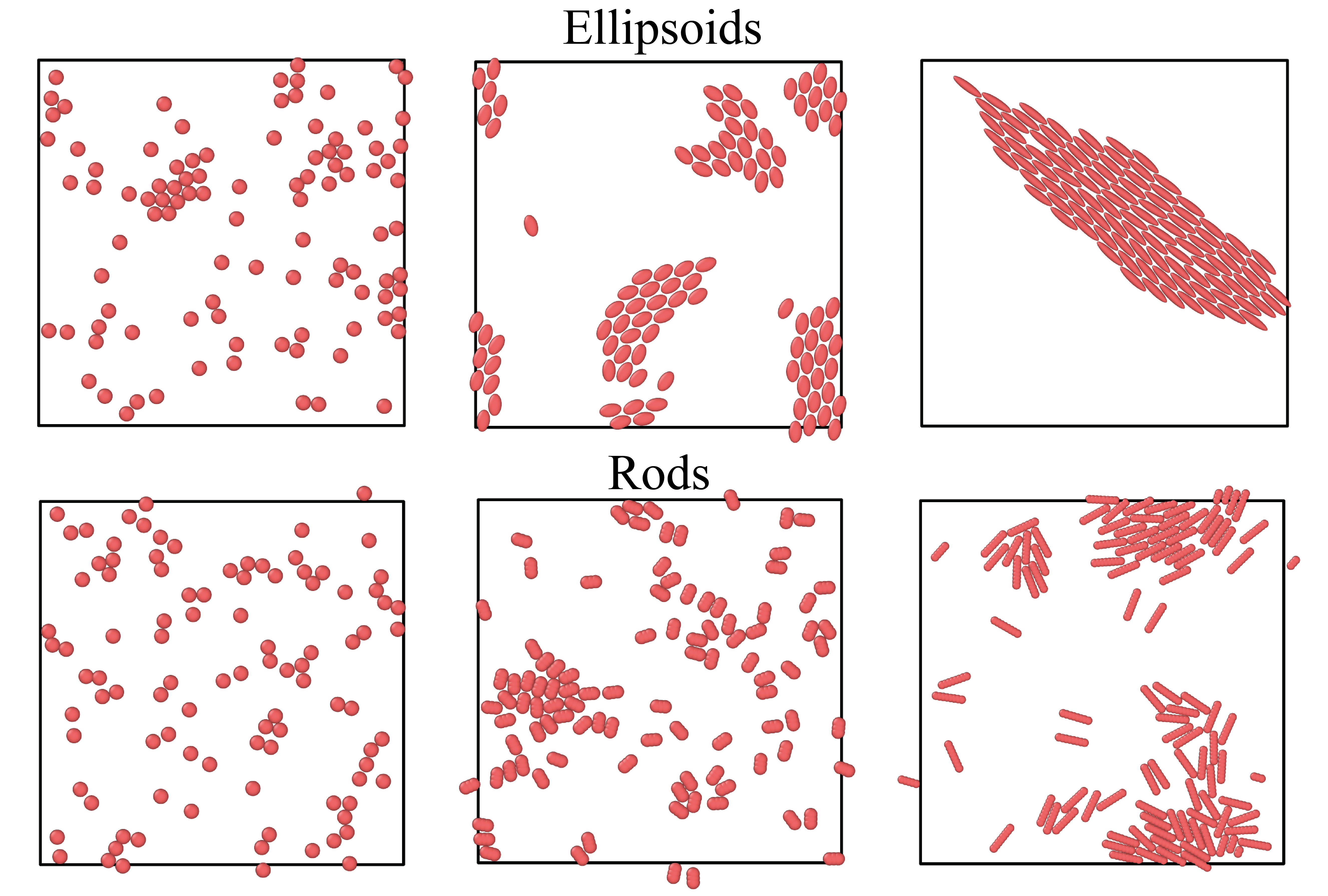}%
    \hfill
    \caption{Configurations obtained for simulations in bulk systems, with periodic boundary condition in both directions, for ellipsoids (top row) and rods (bottom row) and, from left to right, of aspect ratio $\kappa =1.0$, $1.3$ and $2.0$. Simulations were carried out for $N=100$ particles in a simulation box with the size $21\times 21$ in units of $d_0$.} \label{fig:no-wall-ellipsoid-and-rod}
\end{figure}

We performed simulations for both ellipsoids with anisotropic attractive interaction given by Gay-Berne potential and rod-like particles composed of repulsive spheres described by truncated Lennard-Jones potential.

\textbf{Ellipsoids interaction.} We consider ellipsoids with the aspect ratio $\kappa$ ranging from 1 to 2, and with the constant projected area $A=\pi d_0^2 /4= \pi ld  /4$. Interactions between two ellipsoids and an ellipsoid and a wall particle are given by the Gay-Berne potential \cite{Berardi1998, Berardi2008} defined as

\begin{eqnarray}\label{eq. Gay-Berne}
    U_{GB}(\textbf{A}_{1},\textbf{A}_{2},\vec{r}_{12}) = && U_{r}(\textbf{A}_{1},\textbf{A}_{2},\vec{r}_{12},\gamma) \cdot   \eta_{12}(\textbf{A}_{1},\textbf{A}_{2},\nu) \cdot \nonumber \\ && \cdot \chi_{12}(\textbf{A}_{1},\textbf{A}_{2},\vec{r}_{12},\mu) , 
    \label{eq.GB}
\end{eqnarray}
where the distance dependent part is given by

\begin{equation}
    U_{r} = 4 \epsilon_{GB} \left[ \left( \dfrac{\sigma}{h_{12} + \gamma \sigma} \right) ^{12} - \left(\dfrac{\sigma}{h_{12} + \gamma \sigma}\right) ^{6}\right],
\end{equation}

\noindent $\textbf{A}_{1}$ and $\textbf{A}_{2}$ are the transformation matrices from the simulation box frame to the body frame and $\vec{r}_{12}$ is the center-to-center vector between particles. $U_{r}$ controls the shifted distance dependent interaction based on the distance of closest approach $h_{12}$ of the two particles, and $\gamma$ is the shift parameter, $\epsilon_{GB}$ is the depth of the minimum of $U_r$, $\sigma$ is the minimum effective particle radius which we set to $d_0$ here. $h_{12}$ is computed using the scheme developed by Perram et al. \cite{Perram1996, Perram1985, Perram1984}, we emphasize that $h_{12}$ depends upon $\textbf{A}_{1}$ and $\textbf{A}_{2}$ and the particles aspect ratios. $\eta_{12}$ and $\chi_{12}$ terms in Eq.~\ref{eq. Gay-Berne} quantify additional orientation and position dependent contributions to the pair interaction energy, as defined by Everears et al. in \cite{Everaers2003}. Finally, $\nu$ and $\mu$ are some empirical exponents, which we set to unity for simplicity. $\chi_{12}$ term also depends on two energy parameters (which we set here to $k_B T$) describing potential well depths for side-to-side, and end-to-end 
particle orientations, see Ref.~\cite{Everaers2003} for more details. The Gay-Berne potential in Eq.~\ref{eq. Gay-Berne} also describes interaction between two unlike ellipsoids \cite{Everaers2003}, and for two spheres one recovers the standard Lennard-Jones potential. We also employ an interaction cut-off centre-to-centre distance $r_{cut}= 3d_0$ which is larger than the maximum particle length $l_{max}= \sqrt{2}d_0$ considered here. Finally, we set $\epsilon_{GB} = 2$ in units of kinetic energy of a single active particle at terminal velocity.

\textbf{Rods interaction.} For rods, we consider the same values of the aspect ratio as for the case ellipsoids. A rod particle with the length $l$ is composed of $n_r$ repulsive spheres of diameter $d$ with an overlap, $\Delta d\geq d/2$.

The interaction between the beads of two different rod is defined by a truncated Lennard Jones potential,

\begin{equation}\label{eq. Lennard-Jones}
	U_{LJ}(r^*) = 4 \epsilon_{LJ} \bigg[\bigg( \dfrac{\sigma^{*}}{r^{*}}\bigg) ^{12} - \bigg( \dfrac{\sigma^{*}}{r^{*}}\bigg) ^{6}\bigg] , \quad r^{*}<r_c,
\end{equation}

\noindent where $r^{*}$ is the center-to-center distance between two beads of different rods, $r_c$ is the cut-off distance, $\epsilon_{LJ}$ is the well-depth parameter and $\sigma^{*}$ is the distance at which the particle-particle potential is zero. We set $\epsilon_{LJ} = 2$, in units of kinetic energy of a single active particle at terminal velocity, $r_c=d$, and $\sigma^{*}$ is determined from the condition $r_{c}=2^{1/6}\sigma^{*}$. The interaction between a rod and a wall particles is also described by Eq.~\ref{eq. Lennard-Jones}.

\begin{figure}[htb]
\includegraphics[width=1.0\columnwidth]{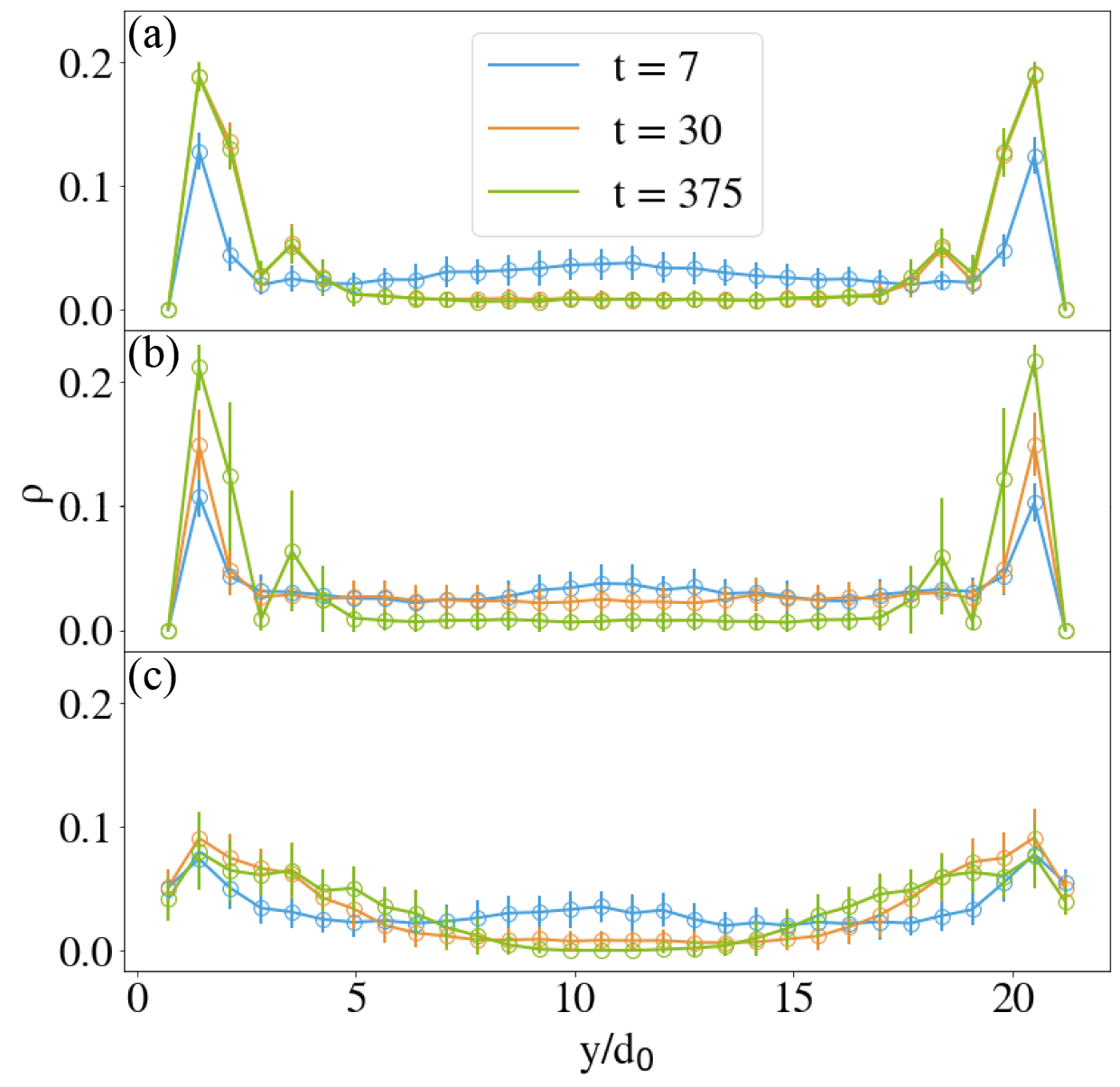}
\caption{Laterally averaged number density $\rho$ of active ellipsoids as a function of the distance $y$ to one of the slit walls
    at several values of time $t$ (given in in units of the ballistic time $\tau_b=d_0/v$) and at (a) $\kappa = 1.0$; 
   (b) $\kappa = 1.2$; (a) $\kappa = 1.9$.  
    The results are obtained at $N=200$, the slit length $L = 42$ and width $H = 21$ units of $d_0$; and are obtained after averaging over 100 independent runs.}
\label{fig:rho-ellipsoid}
\end{figure}

\begin{figure}[htb]
\includegraphics[width=0.8\columnwidth]{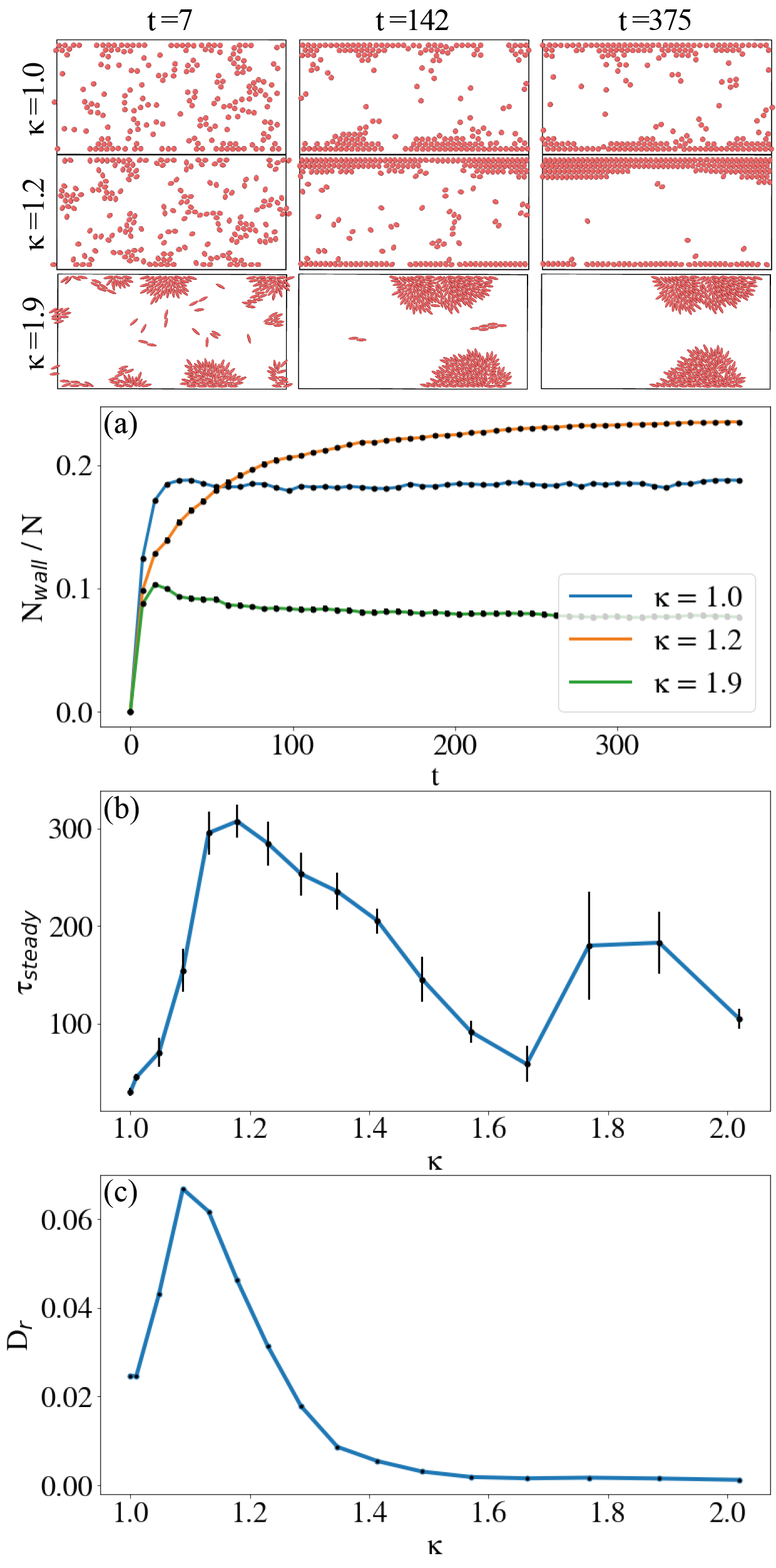}
\caption{
  (a) The fraction $N_{wall}/N$ of particles in a direct contact with the walls as a function of time $t$ (in units of the ballistic time $\tau_b=d_0/v$) for different
    values of the aspect ratio $\kappa = 1.0$, $1.2$ and $1.9$. The simulation snapshots obtained at several $t$ and for these values of $\kappa$
    are shown at the top.
    (b) The relaxation time $\tau_{Steady}$ needed to reach the steady state as a function of $\kappa$.
    (c) Rotational diffusion coefficient as a function of the aspect ratio $\kappa$.
    The results are obtained at $N=200$, the slit length $L = 42$ and width $H = 21$ in units of $d_0$.
    The results in (a), (b), and (c) are obtained after averaging over 100 independent runs. }\label{fig:timeseries-ellipsoid}
\end{figure}

\section{Results}
\label{results}

In Fig.~\ref{fig:no-wall-ellipsoid-and-rod}, we report steady state snapshots of active ellipsoids (top) and rods (bottom) in systems with periodic boundary conditions applied in both directions, for several values of $\kappa$. Both types of active particles form clusters which is reminiscent of the flocking behaviour observed in the Vicsek model \cite{Vicsek2012}. The tendency to form clusters is more pronounced for larger aspect ratios and is additionally enhanced for ellipsoids due to their attractive aligning interaction.

\subsection{Surface aggregation dynamics of active ellipsoids}

Here, we focus on the dynamics of formation of ellipsoid clusters and films near the walls. In Fig.~\ref{fig:rho-ellipsoid} we present the laterally averaged (along the $x-$axis) number density profile $\rho$ as a function of the the distance $y$ to one of the walls and at several values of time $t$ and aspect ratio $\kappa$. The main characteristic shared by all the $\rho(y)$ curves is the accumulation of the active particles at the walls. This is in agreement with the earlier studies \cite{Wensink2008-aggregation-colloidal-rods-near-walls, Elgeti2009-rods-near-surface}. For small values of $\kappa$ the surface density $\rho(y/d_0=2,t)$ is a monotonous function of $t$, showing faster approach to the steady state for smaller values of $\kappa$ (compare Fig.~\ref{fig:rho-ellipsoid}(a) with Fig.~\ref{fig:rho-ellipsoid}(b)). Surprisingly, for larger values of $\kappa$, $\rho(y/d_0=2,t)$ behaves in a non-monotonic way with $t$ (see Fig.~\ref{fig:rho-ellipsoid}(c)). As we discuss below, this behavior can be explained by initial adsorption of small clusters which at later times coalesce resulting in a decrease of the wall contact density.

\begin{figure}[htb]
    \includegraphics[width=1.0\columnwidth]{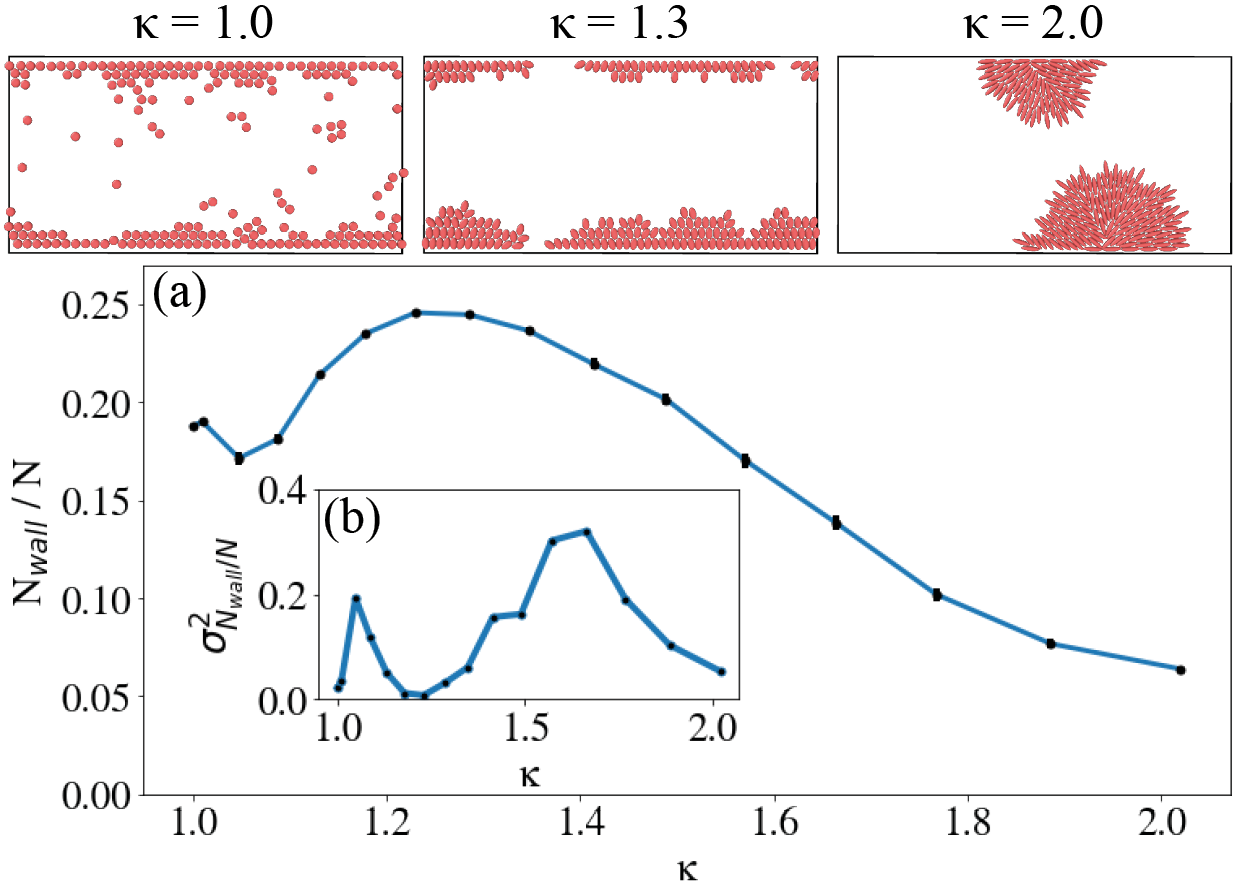}%
    \hfill
    \caption{Typical snapshots of to the steady states of active ellipsoids at different values of aspect ratio $\kappa$ = 1.0, $\kappa$ = 1.3 and $\kappa$ = 2.0.
    (top).
      (a) The fraction $N_{wall}/N$ of the ellipsoids in contact with the walls at the steady states as a function of
        $\kappa$. (b) The variance, $\sigma^2_{N_{wall}/N}$, of $N_{wall}/N$ as a function of $\kappa$. The results are obtained at $N=200$,
        the slit length $L = 42$ and width $H = 21$ in units of $d_0$. The results in (a) and  (b) are obtained
        after averaging over 100 independent runs.}
    \label{fig:fraction-ellipsoids}
\end{figure}


Figure \ref{fig:timeseries-ellipsoid} (a) shows the temporal evolution of the number of particles $N_{wall}$ in a contact with the walls for different aspect ratios of the ellipsoids. The selection criterion for a particle to be in a contact with a wall is based on the "surface-to-surface" distance between an ellipsoid and a wall which must be $\leq d/2$. For small $\kappa$, we find a rapid evolution towards a steady state, which indicates that active particles close to spherical shape form stable surface films faster (see Fig.~\ref{fig:timeseries-ellipsoid}(b)). However, for larger aspect ratios the approach to steady state is slower, see orange curve in Fig.~\ref{fig:timeseries-ellipsoid}(a) with $\kappa = 1.2$. Next, we extract from the curves $N_{wall}(t)$ the time $\tau_{Steady}$ required for $N_{wall}$ to saturate. Interestingly, $\tau_{Steady}$ exhibits a non-monotonic behavior as a function of $\kappa$, as shown in Fig.~\ref{fig:timeseries-ellipsoid}(b). Initially, for $\kappa \geq 1$, $\tau_{Steady}$ increases rapidly and reaches the global maximum at $\kappa \approx 1.2$, followed by a decrease to a local minimum at $\kappa \approx 1.7$. The emergence of the global maximum can be related to a counter-intuitive non-monotonic dependence of the coefficient of rotational diffusion $D_r$ on the aspect ratio. Recall that for passive particles it is expected that $D_r$ decreases monotonously with the increase of $\kappa$. In the case of active ellipsoids, we find that $D_r$ increases rapidly for small $\kappa$ and attains its maximum at $\kappa \approx 1.1$, see Fig.~\ref{fig:timeseries-ellipsoid}(c), which we attribute to a collision-induced enhancement of the particle reorientation. Eventually, $D_r \approx 0$ for $\kappa \gtrsim 1.6$ when the attractive aligning interaction leads to the  ellipsoid clustering and to suppression of the particle angular fluctuations.

The adsorption of small aggregates at the wall (shown in the third row of Fig.~\ref{fig:timeseries-ellipsoid}, $\kappa = 1.9$) can qualitatively explain
 the existence of a peak in the ${N_{wall}}(t,\kappa = 1.9)$ visible at early times in Fig.~\ref{fig:timeseries-ellipsoid}(a).
 The initial rapid increase of ${N_{wall}}$ is due to the adsorption of small flocks, which at later times tend to coalesce to form larger clusters
 (central and right panels of Fig.~\ref{fig:timeseries-ellipsoid} at $\kappa = 1.9$). The coalescence effectively reduces ${N_{wall}}$ and increases the relaxation time
 $\tau_{Steady}$ due to the low mobility of wall-bounded clusters. This explains qualitatively the existence of a local maximum in the  $\tau_{Steady}(\kappa)$
 curve at $\kappa \approx 1.9$ (Fig.~\ref{fig:timeseries-ellipsoid}(b)) since ellipsoids with larger $\kappa$ need more time to form aggregates and to reach the steady state. Similar non-monotonic behavior is revealed by the near surface number density $\rho(y/d_0=2,t)$ as a function of time $t$, see Fig.~\ref{fig:rho-ellipsoid}(c). Additionally, active ellipsoids exhibit flocking behavior for large enough $\kappa$ as demonstrated in Fig.~\ref{fig:timeseries-ellipsoid} at $\kappa = 1.9$.
  The formation of the orientationally ordered flocks is facilitated by the aligning Gay-Berne interaction. As the result the clusters 
  are very mobile while away form the walls. In contrast,  the clusters adsorbed at the walls are much slower which is related to a hedgehog-like alignment of the ellipsoids in such
 clusters (Fig.~\ref{fig:timeseries-ellipsoid}, $\kappa = 1.9$), similar behavior was also reported in \cite{Wensink2008-aggregation-colloidal-rods-near-walls}.


\subsection{Steady-state surface aggregates of ellipsoids}

 Next we discuss how the structure of steady states of active ellipsoids depends on the aspect ratio, which is summarized in Fig.~\ref{fig:fraction-ellipsoids}(a).
   In particular, we focus on the dependence of $N_{wall}$ on $\kappa$ which reveals a local minimum at small $\kappa$, a maximum at intermediate aspect ratios, and a monotonous decrease for larger $\kappa$. The existence of the local minimum
   can be rationalized in terms of the duration of pairwise collisions, which are know to be sensitive to torques operating between elongated particles
   \cite{VanDamme2019-interparticle-torques-suppress-MIPS}. Thus, for hard active rods the excluded volume torques reduce the collision time, which we assume is also the case
   for our active Gay-Berne ellipsoids with not too large aspect ratio. In turn, the shorter collision time suppresses the tendency of ellipsoid to form clusters thereby
   reducing ${N_{wall}/N}$ at the steady state. These qualitative arguments are supported quantitatively by the behavior of the rotational diffusion coefficient $D_r$, which has a maximum at $\kappa \approx 1.1$, see Fig.~\ref{fig:timeseries-ellipsoid}(d). Larger $D_r$ results in faster particle reorientation, which suppresses the tendency to aggregate at the walls.

   For longer ellipsoids the effect of the excluded volume torques will be less pronounced due to the increased role of the attractive orientational interaction
   leading to the formation of flocks which subsequently accumulate at the walls. This leads to an increase of $N_{wall}/N$ for $1.1 \lesssim \kappa \lesssim 1.2$ followed by a decrease for larger values of $\kappa$. This decrease indicates a configurational transition from layered structures of ordered particles observed for intermediate values of $\kappa$ (see middle snapshot in Fig.~\ref{fig:fraction-ellipsoids} at $\kappa = 1.3$) to hedgehog-like clusters for larger ones (see snapshot in Fig.~\ref{fig:fraction-ellipsoids} at $\kappa = 1.9$). This transition is also reflected in the dependence of the variance $\sigma^2_{N_{wall}/N}$ of $N_{wall}/N$ on the aspect ratio, with a characteristic peak at $\kappa \approx 1.6$ depicted in Fig.~\ref{fig:fraction-ellipsoids}(b).

\begin{figure}
\includegraphics[width=1.0\columnwidth]{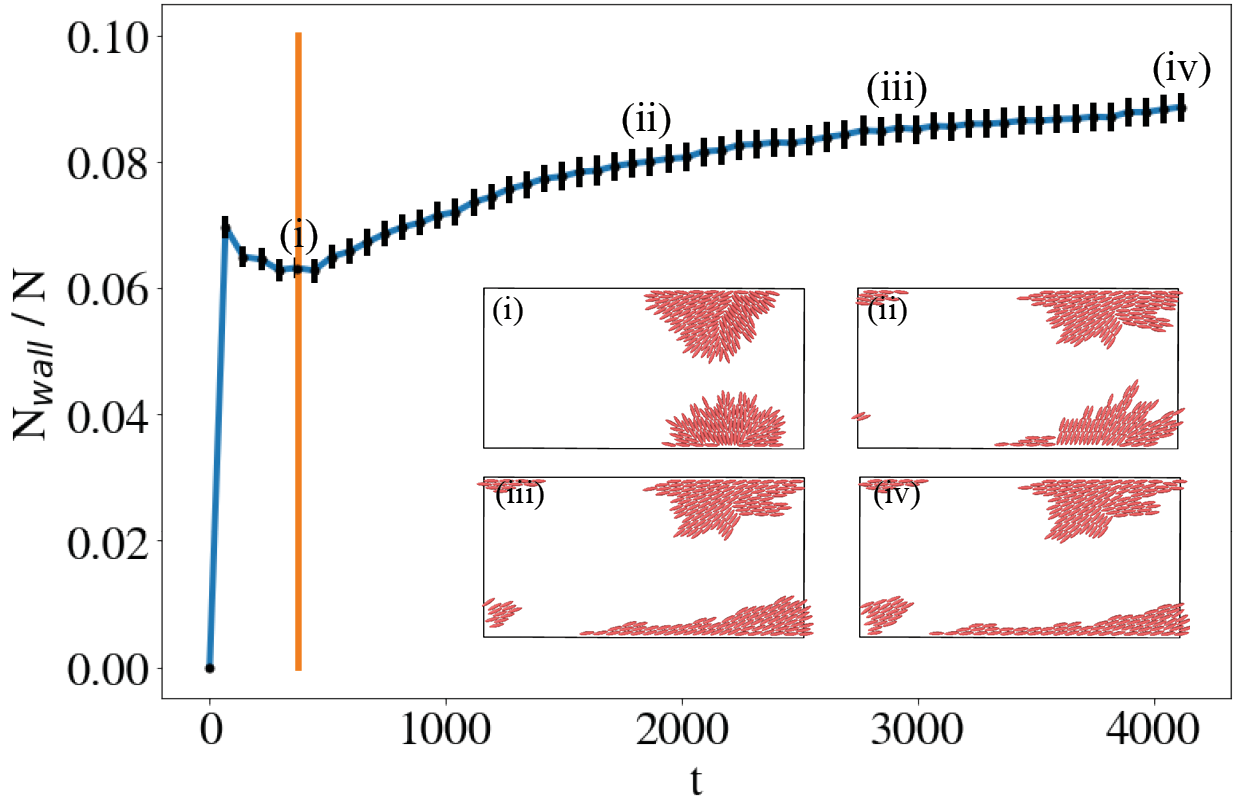}
\caption{ (a) The fraction $N_{wall}/N$ of the ellipsoids in a direct contact with the walls as a function of time $t$, at $\kappa = 2.0$. The particle activity is switched off at $t = 375\tau_b$. (i)-(iv) the simulation
    snapshots obtained at times marked along the $N_{wall}(t)/N$ curve in (a). $N=200$, the slit length $L = 42$ and width $H = 21$ in units of $d_0$. The results are obtained after averaging over 50 independent runs.}
\label{fig:no-activity-ellipsoid}
\end{figure}

In order to test which of the two factors, i) the particle activity or ii) the attractive orientational interaction, are more important in the formation of the hedgehog-like wall structures, we switched off the particle activity at some instance of time and monitor the subsequent system evolution. Fig.~\ref{fig:no-activity-ellipsoid} shows $N_{wall}(t)/N$ calculated for this
situation at $\kappa = 2.0$; several representative snapshots are demonstrated as well. The suppression of the particle activity results in the ``melting'' of the hedgehog-like ordering of the particles and a subsequent spreading of the clusters along the walls.


\begin{figure}[htb]
    \includegraphics[width=0.95\columnwidth]{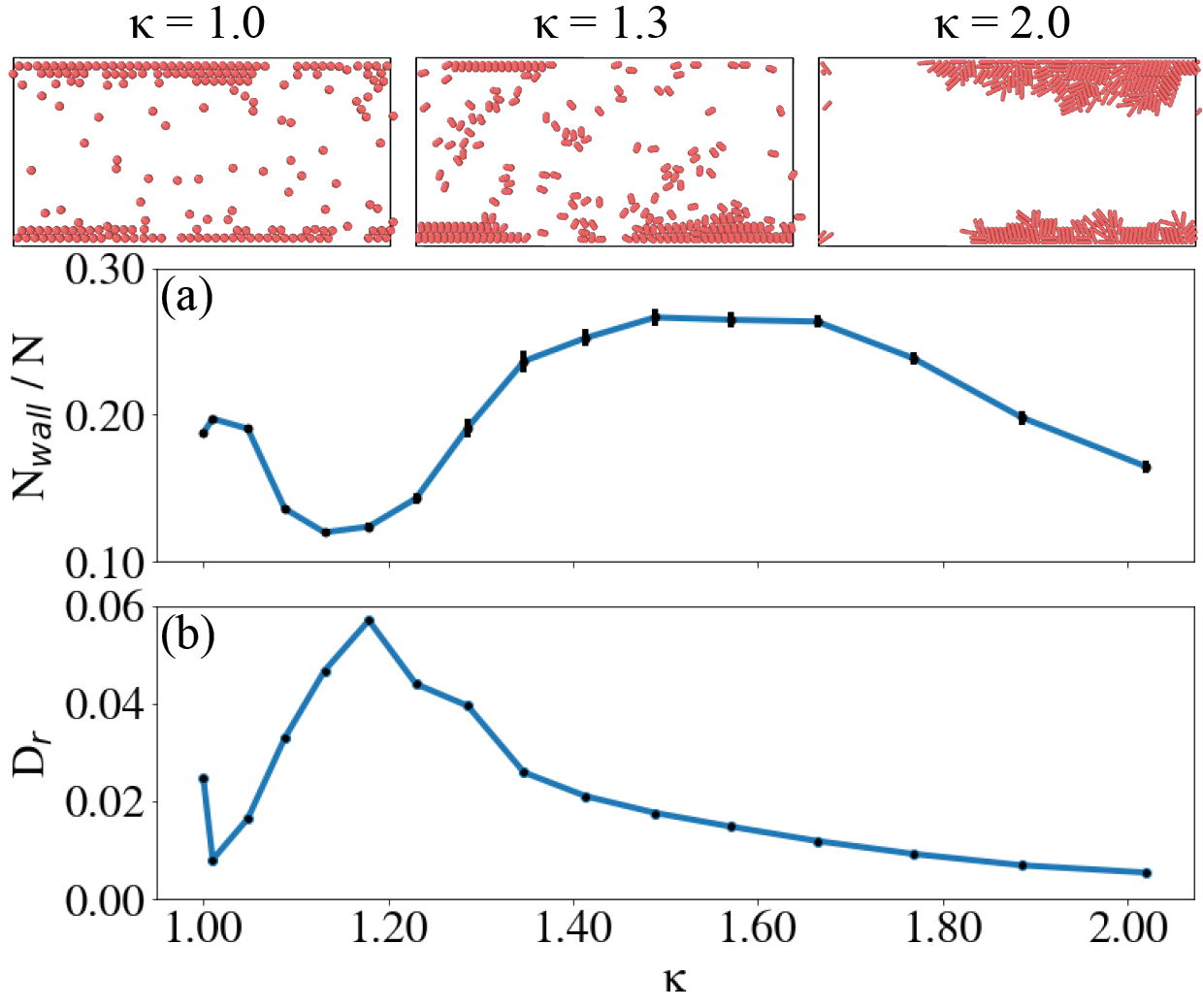}
    \caption{ Typical snapshots corresponding to the steady states of active rods at different values of the aspect ratio $\kappa = 1.0, 1.3, 2.0$ (top).
       (a) The fraction $N_{wall}/N$ of the ellipsoids in contact with the walls at the steady state as a function of $\kappa$. 
       (b) Rotational diffusion coefficient $D_r$ of active rods as a function of aspect ratio $\kappa$.
       The results are obtained at $N=200$,the slit length $L = 42$ and width $H = 21$ in units of $d_0$. The results in (a) and  (b) are obtained
        after averaging over 100 independent runs.}\label{fig:fraction-rods}
\end{figure}

\subsection{Steady-state surface aggregates of active rods}

We performed similar study for active rods with repulsive interaction only, in order to understand the effect of the Gay-Berne potential in the previous section. The snapshots at the top of Figure \ref{fig:fraction-rods} show a similar tendency to form compact clusters near the walls for large aspect ratios. However these aggregates are not as stable as the the ones formed by the ellipsoids, and have a tendency to appear and disappear. Additionally, the surface aggregates of the rods tend to form film like structures at late times, see top-right panel in Fig.~\ref{fig:fraction-rods} corresponding to $\kappa = 2.0$. We attribute this difference between the clusters of ellipsoids and rods to the lack of orientational attraction between the active rods. ${N_{wall}/N}$ as a function of $\kappa$ demonstrate similar trends to what was described above for the case of the ellipsoids (compare with Fig.~\ref{fig:fraction-ellipsoids}(a)). In the present case we find a local minimum at $\kappa \approx 1.15$, which however is much more pronounced compared to the previous case. Now, the absence of attractive orientational interactions translates in a larger range of $\kappa$ where the torques (excluded volume) shorten the duration of the collisions and suppress the process of aggregation at the wall. The tendency to form clusters increases with the increase of $\kappa$. We also verified that the rotational diffusion coefficient $D_r$ for the rods exhibit qualitatively similar dependence on $\kappa$ (see Fig.~\ref{fig:fraction-rods}(b)).

    \section{Conclusions}
\label{conclusions}

In this paper we show how anisotropic interactions, known to change the universality class of a growing interface of passive colloids \cite{Yunker2013, Dias2018, Dias2014, Araujo2015}, affect the formation of structures of active particle near walls. Taking into account the excluded volume torques only, as in the of the  case rod-like particles, we observe similar steady states configurations at the wall as those previously shown by Wensink et al. \cite{Wensink2008-aggregation-colloidal-rods-near-walls}, where short-living hedgehog-like clusters were reported. However, when including the anisotropic attractive interactions as described by Gay-Berne potential, we find the formation of long-living hedgehog-like clusters of active ellipsoids near the wall. A configurational transition form layered surface aggregates to the formation of the hedgehogs is observed at the aspect ratio threshold $\kappa \approx 1.6$.

We also found that the previously proposed explanation for the suppression of MIPS for rod-like particles \cite{VanDamme2019-interparticle-torques-suppress-MIPS}, affects cluster formation near the wall, where ellipsoids with smaller aspect ratio tend to have a slower evolution to a steady state due to collisions with the boundary and with other particles that increase the rotational diffusion coefficient. This effect diminishes with the increase of the aspect ratio due to a drastic decrease of the rotational diffusion coefficient, which in turn enhances the  aggregation at the walls. For ellipsoids with the aspect ratios around 2, small clusters initially form near the walls, which  at later times come together to form  bigger clusters, occupying a smaller portion of the walls. This coalescence dynamics at the surfaces slows down the approach to a steady state. 

The dynamics of anisotropic active matter near walls can impact interfacial phenomena in many problems related to growing films. For instance, the structure of interfaces  of active nematic has shown a dependence on the anchoring or friction with a substrate \cite{Coelho2020, Coelho2021} which influences vortex formation and the behavior of sessile nematic drops. Also the presence of passive obstacles at a wall can affect clustering dynamics, since it was shown that the interaction of active particles with passive ones can enhance the propagation of the active particles along the surface \cite{Reichhardt2014, Volpe2017,Makarchuk2019}. For active rods moving in the presence of fixed surface obstacles, an optimal density of obstacles was predicted, which enhances the rods diffusivity for large enough aspect ratio \cite{Khalilian2016}, and when active Brownian particles experience a short-range aligning interaction, they form a network of tunnels in the presence of passive particles \cite{Nilsson2017}. Finally, for cellular systems, it is known that the surrounding conditions affect cell motility \cite{Paul2016, Pinto2020,Dias-colloid_scaffolds-2020} where both one-dimensional confinement \cite{Pags2020} or two-dimensional substrates can increase cell motility \cite{Melo2021}.

\begin{acknowledgments}
We acknowledge financial support from the Portuguese Foundation for Science and Technology (FCT) under Contracts no. PTDC/FIS-MAC/28146/2017 (LISBOA-01-0145-FEDER-028146), PTDC/FIS-MAC/5689/2020, CEECIND/00586/2017, UIDB/00618/2020,  and UIDP/00618/2020.  
\end{acknowledgments}

\bibliography{apssamp}

\end{document}